\documentclass[12pt,journal,a4paper,onecolumn]{IEEEtran}

\usepackage{amsopn,amsmath,amssymb,amsfonts,bm,amsthm}
\usepackage{graphicx,subfigure}
\usepackage{enumerate}
\usepackage{cite}
\usepackage{etex}
\usepackage{algorithm,algorithmic}
\usepackage{tikz,pgfplots}
\usepackage[margin=16mm,top=20mm,bottom=16mm]{geometry}

\usetikzlibrary{plotmarks}
\pgfplotsset{compat=1.5}
\IEEEoverridecommandlockouts

\def\mA{\boldsymbol A}
\def\mG{\boldsymbol G}
\def\PP{\mathbf p}

\def\E{\mathcal E}
\def\G{\mathcal G}
\def\cG{\overline{\G}}
\def\M{\mathcal M}
\def\P{\mathcal P}
\def\S{\mathcal S}
\def\T{\mathcal T}
\def\V{\mathcal V}
\def\W{\mathcal W}
\def\X{\boldsymbol X}
\def\F{\mathbb F}
\def\p{\mathbf p}

\def\wmax{W_{\max}}
\def\UIDNC{U_{\textrm{IDNC}}}
\def\URLNC{U_{\textrm{RLNC}}}
\def\DIDNC{D_{\textrm{IDNC}}}
\def\DRLNC{D_{\textrm{RLNC}}}

\newcommand{\secref}[1]{Section\,\ref{#1}}

\newcommand{\figref}[1]{Fig.\,\ref{#1}}

\newtheorem{Lemma}{\textbf{Lemma}}

\newtheorem{Corollary}{Corollary}
\newtheorem{Definition}{\textbf{Definition}}

\newtheorem{Remark}{\textbf{Remark}}

\newtheorem{Example}{\textbf{Example}}
  {\proof}{\proofend}
\newtheorem{Proposition}{\textbf{Proposition}}

\author{
  \IEEEauthorblockA{Mingchao Yu~~~~~~~~~~~~~Neda Aboutorab~~~~~~~~~~~Parastoo Sadeghi}\\
  \IEEEauthorblockA{\small Research School of Engineering, The Australian National University, Canberra, Australia \\ \texttt{\{ming.yu, neda.aboutorab, parastoo.sadeghi\}@anu.edu.au}}}
\title{From Instantly Decodable to Random Linear Network Coding}
\begin{document}
\sloppy

\maketitle

\begin{abstract}
Our primary goal in this paper is to traverse the performance gap between two linear network coding schemes: random linear network coding (RLNC) and instantly decodable network coding (IDNC) in terms of throughput and decoding delay. We first redefine the concept of packet generation and use it to partition a block of partially-received data packets in a novel way, based on the coding sets in an IDNC solution. By varying the generation size, we obtain a general coding framework which consists of a series of coding schemes, with RLNC and IDNC identified as two extreme cases. We then prove that the throughput and decoding delay performance of all coding schemes in this coding framework are bounded between the performance of RLNC and IDNC and hence throughput-delay tradeoff becomes possible. We also propose implementations of this coding framework to further improve its throughput and decoding delay performance, to manage feedback frequency and coding complexity, or to achieve in-block performance adaption. Extensive simulations are then provided to verify the performance of the proposed coding schemes and their implementations.
\end{abstract}

\section{Introduction}\label{sec:intro}
Network coding allows senders or intermediate nodes of a
network to mix different data packets/flows and can enhance
the throughput of many network setups
\cite{Yeung_flow,koetter:medard:2003}. But this is often at
the price of large decoding delay, because mixed data needs to be
network decoded before delivery to higher layers
\cite{fragouli:lun:medard:pakzad:2007,eryilmaz:ozdaglar:medard:ahmed:2008}.
Understanding the tradeoff between throughput and
decoding delay in network coded systems has been the
subject of research in recent years
\cite{fragouli:lun:medard:pakzad:2007,eryilmaz:ozdaglar:medard:ahmed:2008,costa:munaretto:widmer:baros:2008,swapna:eryilmaz:shroff:2010},
where a packet-level network coding model is particularly
suitable for such studies. In this paper, we are primarily concerned with the achievable throughput-delay tradeoff of packet-level network coding in wireless
broadcast scenarios, where a sender wishes to broadcast a block of data packets to some receivers through wireless channels with packet erasures. We first review two
classic packet-level network coding techniques under this scenario: random linear network coding (RLNC)
\cite{ho:medard:koetter:karger:effros:2006} and instantly
decodable network coding (IDNC)
\cite{sadeghi:shams:traskov:2010,sorour:valaee:2010,yu:parastoo:neda:idnc2013}.

The primary advantage of RLNC is its optimality in terms of block completion time
\cite{eryilmaz:ozdaglar:medard:ahmed:2008}, which is the time it takes to successfully broadcast all data packets in a block to all receivers and is a fundamental measure of throughput. RLNC achieves its optimality by sending random linear combinations of all
data packets as coded packets, with coefficients randomly chosen from a
finite field $\F_q$ where usually $q\gg2$.  Another advantage of RLNC is that it only
requires one ACK feedback from each receiver upon
successful decoding of all data packets in a block. However,
RLNC can suffer from large decoding delays, since
a receiver needs to collect enough linear combinations to perform block-wise decoding. This also incurs heavy decoding computational load, since Gaussian eliminations on a coefficient matrix under large field size $q$ are involved \cite{ho:medard:koetter:karger:effros:2006}.

On the contrary, IDNC aims at minimizing
decoding delay. By carefully choosing data packets
to be coded together, it guarantees a subset of (or if
possible, all) receivers to instantly decode one of
their wanted data packets upon successful reception of a
coded packet
\cite{sadeghi:shams:traskov:2010,yu:parastoo:neda:idnc2013}. This
means that data packets can be potentially delivered to higher layers much faster than RLNC. Another advantage of IDNC is its simple XOR-based encoding and decoding,
i.e., coding coefficients are chosen from the binary field $\F_2$. However,
IDNC is generally not optimal in terms of block completion
time, as there might exist a subset of receivers who cannot instantly decode any of their wanted data packets from the received coded packet. Moreover, IDNC requires feedback from all
receivers at a proper frequency for making coding decisions.

\section{Problem Description and Contributions}\label{sec:contributions}
RLNC and IDNC have many contrasting features. They generally trade off decoding delay or throughput for one another, respectively. They also differ in practical issues such as feedback frequency and decoding complexity. Moreover, since RLNC uses a larger finite field than IDNC, any adaptive choice and switching between IDNC and RLNC is impossible during the broadcast of one block.

In this paper, we are interested in investigating the
performance spectrum between IDNC and RLNC. This study will help us develop coding schemes that offer moderate throughput-delay tradeoffs, or equivalently, more balanced throughput and decoding delay performance compared with RLNC and IDNC. We are also interested in the connection between the coding mechanisms of IDNC and RLNC. This study will help us manage feedback frequency, decoding complexity, and even in-block performance adaption.

The first question we ask is:
\begin{itemize}
\item \emph{Is there a coding scheme which provides more balanced throughput and decoding delay performance than IDNC and RLNC?}
\end{itemize}

At the beginning of \secref{sec:partition_motivation},
we will show the existence of such a scheme through an example, in which the decoding delay can be reduced without trading off the throughput. The key idea is to appropriately partition a partially-received block of data packets into \emph{sub-generations} based on their reception states at the receivers after sending them uncoded once first. A similar partitioning has been studied in the literature \cite{emina:joshi:isit2013}, where sub-generations with equal number of data packets are generated and transmitted separately to reduce decoding delay in peer-to-peer scenarios. However, as we will prove in \secref{sec:partition_motivation}, such partitioning may generally yield a prohibitively large block
completion time compared with RLNC and IDNC in our broadcast scenario. Hence, the following question arises:
\begin{itemize}
\item \emph{Is there a better way of packet partitioning to achieve more balanced throughput and decoding delay performance during block transmission?}
\end{itemize}

Our answer to this question is through using the concept
of \emph{non-conflicting data packets}
\cite{yu:parastoo:neda:idnc2013}, which is defined as data packets not jointly wanted by any receiver. By properly grouping non-conflicting data packets together into coding sets, we obtain an IDNC solution \cite{yu:parastoo:neda:idnc2013}. Then by appropriately partitioning coding sets in an IDNC solution into sub-generations, we obtain
a desirable balance between throughput and decoding delay performance, which lies between that of RLNC and IDNC. Consequently, we redefine the concepts
of sub-generation and its size, which is the core of this work. Based on these new concepts, we achieve the following contributions\footnote{Preliminary results of our work have been partly published in \cite{yu:neda:parastoo:isit2013}.}:
\begin{enumerate}[(1)]
\item We fill the throughput and decoding delay performance gap between IDNC and RLNC by proposing new adaptive coding schemes. They offer more balanced throughout and decoding delay performance than IDNC and RLNC;
\item The new coding schemes, together with IDNC and RLNC, constitute a general coding framework. This coding framework unifies IDNC and RLNC, because they become its two specific cases;
\item We study the throughput and decoding delay properties of this coding framework. The results provide a good insight from the perspective of the proposed sub-generation size;
\item We develop various implementations of this coding framework, which can further improve throughput and decoding delay performance without sacrificing on one another. They also enable in-block performance adaption by allowing switching between IDNC and RLNC.
\end{enumerate}

Plenty of hands-on examples are designed to demonstrate the proposed concepts and algorithms. Extensive simulations are also provided to evaluate the performance of the proposed coding schemes.

\section{System Model}\label{sec:system}

\subsection{Transmission Setup}
We consider wireless broadcast of $K_T$ data packets, denoted by $\p_1,\cdots,\p_{K_T}$, from a sender to $N_T$ receivers, denoted by $R_1,\cdots,R_{N_T}$. Time is slotted. In each time slot, a packet (either original data or coded) is sent. Wireless channels between the sender and the receivers are subject to packet erasures. For the ease of simulations, we assume i.i.d. memoryless erasure channels to all receivers with equal probability $P_e$. However, this is not strictly needed and can be easily extended to more general cases.

In this scenario, $K_T$ data packets are first sent uncoded once using $K_T$ time slots. This phase is known as a \emph{systematic transmission phase} \cite{sadeghi:shams:traskov:2010,keller:drinea:fragouli:2008,heide_systematic_RLNC}. We justify the application of this phase on IDNC and RLNC in the following Remark. We will review this again under the proposed coding framework.

\begin{Remark}\label{remark:systematic}
For IDNC, a systematic transmission phase is compulsory \cite{sundararajan:sadeghi:medard:2009}, since coding opportunities are unavailable at the beginning. RLNC with a systematic transmission phase is known as systematic RLNC \cite{keller:drinea:fragouli:2008,heide_systematic_RLNC}. It reserves the throughput optimality of the traditional RLNC. It also introduces extra benefits such as 1) smaller packet decoding delay, and 2) smaller field size $q$ and thus lower coding/decoding complexity.
In the rest of this paper, the term RLNC always refers to systematic RLNC unless otherwise specified.
\end{Remark}

\begin{figure}
\centering
\subfigure[SFM $\mA$]{\includegraphics[width=0.45\linewidth]{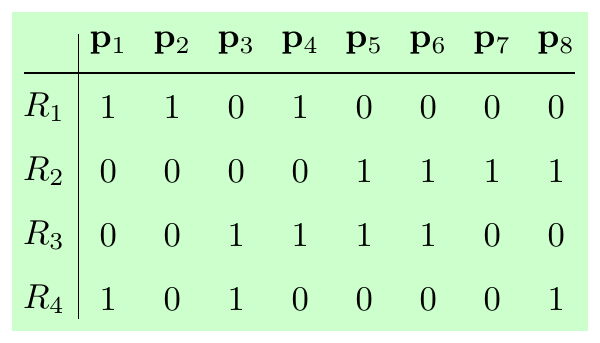}\label{fig:sfm}}\hspace{25pt}
\subfigure[Graph $\G$]{\includegraphics[width=0.2\linewidth]{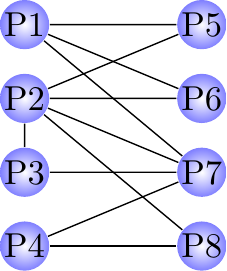}\label{fig:graph}}
\caption{An example of the state feedback matrix $\mA$ and its IDNC graph.}\label{fig:SFM_example}
\vspace{-1em}
\end{figure}

During the systematic transmission phase a receiver might miss any data packet due to packet erasures. The receivers then send lossless feedback to the sender about their packet reception states. Feedback information can be expressed by an $N\times K$ binary state feedback matrix (SFM) \cite{sadeghi:shams:traskov:2010,sameh:valaee:globecom:2010}, denoted by $\mA=[a_{n,k}]$, where $a_{n,k}=1$ means receiver $R_n$ still wants data packet $\p_k$ and $a_{n,k}=0$ otherwise. Here $N$ is the number of receivers who have not received all $K_T$ data packets and $K$ is the number of data packets that have not been received by all $N_T$ receivers. The set of these $K$ partially-received data packets is denoted by $\P_K$.  An example of a $4\times8$ SFM is given in Fig. \ref{fig:sfm}. The subset of $\P_K$ wanted by receiver $R_n$ is called the \emph{Wants} set of this receiver, denoted by $\W_n$. Its size is denoted by $W_n$ and the largest $W_n$ across all receivers is denoted by $\wmax$. The subset of receivers that want $\p_k$ is called the \emph{Target} set of $\p_k$, denoted by $\T_k$. Its size is denoted by $T_k$.

\begin{figure}
\centering
\includegraphics[width=0.7\linewidth]{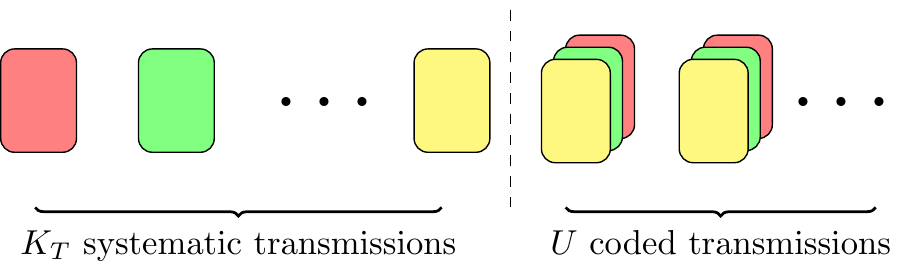}
\caption{The systematic and coded transmission phases.}
\label{fig:phases}
\end{figure}

Based on the SFM, the sender initiates a \emph {coded transmission phase} which is also subject to erasures. In this phase, the sender sends network coded packets to efficiently complete the broadcast of the data block. This two-phase broadcast is illustrated in Fig. \ref{fig:phases}.

The throughput and decoding delay performance of the network coding schemes can be measured by the minimum block completion time and minimum average packet decoding delay, respectively. We first define the minimum block completion time:
\begin{Definition}
Given an SFM and a certain coding scheme, the minimum block completion time, or equivalently the minimum number of coded transmissions, is the smallest possible number of coded transmissions by the sender that is needed to satisfy the demands of all receivers in the absence of any packet erasures in the coded transmission phase. This number is denoted by $U$.
\end{Definition}

The minimum block completion time $U$ reflects the best throughput performance of a certain coding scheme, calculated as $K_T/(K_T+U)$. Such measure of throughput is important in the sense that it disentangles the effect of channel-induced packet erasures and algorithm-induced network coded packet design on the throughput. In reality, due to packet erasures or suboptimal choices of coded packet transmissions, we may need more than $U$ transmissions to complete a data block. In this case, $K_T/(K_T+U)$ serves as an upper bound on throughput and will still serve as a useful measure of performance. We now define the minimum average packet decoding delay:
\begin{Definition}
Given an SFM and a certain coding scheme, assume that a sequence of coded packets that can achieve the corresponding minimum block completion time are transmitted without erasures. Also assume that coded packets are transmitted in the decreasing order of their targeted receivers to minimize decoding delay. Denote by $u_{n,k}$ the first time slot when original data packet $\p_k$ can be decoded by receiver $R_n$, and let $u_{n,k}=0$ if $a_{n,k}=0$. We have $u_{n,k}\in[0,U]$. Then the minimum average packet decoding delay, $D$, is defined as:
\begin{equation}
D\triangleq\frac{1}{\sum_{k=1}^K T_k}\sum_{n=1}^N\sum_{k=1}^Ku_{n,k}\label{eq:d_definition}
\end{equation}
\end{Definition}

It is noted that $D$ refers to the minimum possible delay under a certain coding scheme achieving its minimum block completion time, but not refer to \emph{universal minimum delay} under all coding schemes, which is still an opten question to the best of our knowledge. We now review IDNC and RLNC schemes in the coded transmission phase and discuss their throughput and decoding delay performance.

\subsection{Coded Transmissions Using IDNC}\label{sec:system_idnc}

An IDNC coded packet takes the form of
\begin{equation}\label{eq:idnc:coding}
\X_u = \sum_{\p_k\in\P_k}\beta_{k,u} \p_k
\end{equation}
where $u$ is the time index, $\beta_{k,u} \in \{0,1\}$ and the summation is bit-wise XOR $\oplus$. We denote by $\M_u$ the set of original data packets that have non-zero coefficients in $\X_u$, namely, $\M_u=\{\p_k:\beta_{k,u}=1\}$. $\M_u$ fully represents $\X_u$ and is called a coding set. It is instantly decodable if it satisfies the following \emph{IDNC constraint} \cite{yu:parastoo:neda:idnc2013,sadeghi:shams:traskov:2010}:\footnote{There is another type of IDNC called general IDNC (G-IDNC) \cite{sameh:valaee:globecom:2010,neda:parastoo:o2idnc}, which treats a data packet wanted by several receivers as different data packets and allows transmissions of non-instantly decodable coded packets for a subset of receivers. We do not consider G-IDNC in this work.}

\begin{Definition}\label{def:coding_set}
An IDNC coding set $\M_u$ contains at most one data packet from the Wants set, $\W_n$, of any receiver $R_n$.
\end{Definition}

According to this constraint, we have two important concepts called conflicting and non-conflicting data packets \cite{yu:parastoo:neda:idnc2013}:

\begin{Definition}\label{def:conf_packet}
We say that two data packets $\p_i$ and $\p_j$ conflict with each other if there exists at least one receiver who wants both packets. That is, $\p_i$ and $\p_j$ conflict if both belong to the Wants set, $\W_n$, of at least one receiver such as $R_n$. We say that two data packets $\p_i$ and $\p_j$ do not conflict with each other if there is no receiver who wants both packets.
\end{Definition}

It is clear that to avoid non-instantly decodable coded packets, two conflicting data packets $\p_i$ and $\p_j$ cannot be coded together. Conflict states among all $K$ data packets can be represented by an undirected IDNC graph $\G(\V,\E)$, where vertex $v_k\in\V$ represents packet $\p_k$, and edge $e_{i,j}\in\E$ exists if $\p_i$ does not conflict with $\p_j$ \cite{yu:parastoo:neda:idnc2013}. Below is an example:
\begin{Example}
Consider the SFM in Fig. \ref{fig:sfm}. $\p_1$ and $\p_2$ conflict with each other because $R_1$ wants both packets. Hence, in the IDNC graph $\G$ in Fig. \ref{fig:graph}, $\p_1$ and $\p_2$ are not connected. If $\X=\p_1\oplus\p_2$ were sent, $R_1$ would not be able to instantly decode $\p_1$ or $\p_2$.  On the other hand, $\p_1$ and $\p_5$ do not conflict with each other because there is no single receiver who wants both packets. Thus $\p_1$ and $\p_5$ are connected in $\G$. By receiving $\X=\p_1\oplus\p_5$, $R_1$ can instantly decode $\p_1$ as $\p_1=\X\oplus\p_5$, since $R_1$ already has $\p_5$. Similarly, the other three receivers can also instantly decode one data packet.
\end{Example}

Since data packets in the same coding set $\M$ do not conflict with each other, the vertices representing these packets are connected to each other in $\G$. These vertices form a clique\footnote{A clique is a subset of a graph the vertices in which are all connected with each other.} \cite{Graph_theory} of $\G$. Furthermore, a coding set $\M$ is said to be \emph{maximal} if its corresponding clique is maximal, i.e., is not a subset of a larger clique. We then have the following lemma \cite{yu:parastoo:neda:idnc2013}, which identifies the minimum block completion time of IDNC:
\begin{Lemma}
The minimum block completion time of IDNC, denoted by $\UIDNC$, is equal to the chromatic number\footnote{The chromatic number of a graph is the minimum number of colors one can use to color all vertices such that no two adjacent vertices share the same color. \cite{Graph_theory}} of the complementary IDNC graph $\cG$.
\end{Lemma}
Here, $\cG$ denotes the complementary of $\G$, which has the same vertices as $\G$, but with opposite vertex connectivities. We then have the concepts of an IDNC solution and optimal IDNC solution:
\begin{Definition}\label{def:idnc_solution}
An IDNC solution is a collection of IDNC coding sets which satisfy the following two conditions:
1) they jointly cover all $K$ data packets; and
2) The union of any of them is not an IDNC coding set.
\end{Definition}\label{def:optimal_idnc_solution}
While the first condition ensures  completeness of the solution, the second condition prevents redundant IDNC coding sets which can be absorbed into other coding sets to reduce the cardinality of the solution.
\begin{Definition}
An IDNC solution is optimal if:
1) its cardinality is $\UIDNC$; and
2) all its coding sets are maximal.
An optimal IDNC solution is denoted by $\S=\{\M_1,\cdots,\M_{\UIDNC}\}$.
\end{Definition}
The minimum average packet coding delay of the optimal IDNC solution can be computed using \eqref{eq:d_definition} and is denoted by $\DIDNC$. Below is an example.

\begin{Example}\label{exmp:sfm_idnc}
The optimal IDNC solution for the SFM in Fig. \ref{fig:sfm} is $\S=\{\{\p_2,\p_3,\p_7\}$, $\{\p_4$,$\p_8\}$, $\{\p_1,\p_5\}$, $\{\p_1,\p_6\}\}$. Thus $\UIDNC=4$. One can easily verify that every coding set in $\S$ is maximal.  By sending $\X_1=\p_2\oplus\p_3\oplus\p_7$, $\X_2=\p_4\oplus\p_8$, $\X_3=\p_1\oplus\p_5$, and $\X_4=\p_1\oplus\p_6$ using four time slots, the minimum average packet decoding delay is computed using \eqref{eq:d_definition} to be $\DIDNC=2.3$.
\end{Example}

\subsection{Coded Transmissions Using RLNC}
Coded packets in RLNC are random linear combinations of \emph{all} $K$ data packets:
\begin{equation}
\X_u=\sum_{\p_k\in\P_k}\alpha_{k,u}\p_k
\end{equation}
where coding coefficients $\{\alpha_{k,u}\}$ are randomly chosen from a finite field $\F_q$ and usually $q\gg2$. In order to decode $W_n$ data packets, a receiver $R_n$ needs to receive $W_n$ linearly independent coded packets. Hence, to satisfy the demands of all $N$ receivers, at least $\wmax$ linearly independent coded packets need to be broadcast. The minimum block completion time using RLNC is thus equal to $\wmax$ and is denoted by $\URLNC$. The minimum average packet decoding delay of RLNC, denoted by $\DRLNC$, can be calculated as:
\begin{equation}\label{eq:d_rlnc}
\DRLNC=\sum_{n=1}^{N}W_n^2
\end{equation}
where we ignore early decoding chances of RLNC and assume that decoding of $W_n$ wanted packets by receiver $R_n$ is only possible after $W_n$ coded packets are received.

For the SFM given in Fig. \ref{fig:sfm}, if RLNC is applied, $\URLNC=4$ because $\wmax=4$. The minimum average packet decoding delay is $\DRLNC=3.6$, which is almost $60\%$ larger than IDNC with $\DIDNC=2.3$.

\begin{Remark}\label{remark:rlnc_vs_idnc}
RLNC has a better throughput performance than IDNC, i.e., $\URLNC\leqslant\UIDNC$. This relationship is always valid. Actually, $\URLNC$ is the benchmark for any network coding technique under the scenario considered here. On the other hand, IDNC is expected to have better decoding delay performance than RLNC, i.e., $\DIDNC\leqslant\DRLNC$. However, this relationship is not always valid, as we may find some instances of SFM where $\DIDNC>\DRLNC$. Below is one such example. Detailed comparisons can be found in \cite{yu:parastoo:neda:idnc2013}.
\end{Remark}

\begin{Example}
\begin{figure}
\centering
\includegraphics[width=0.4\linewidth]{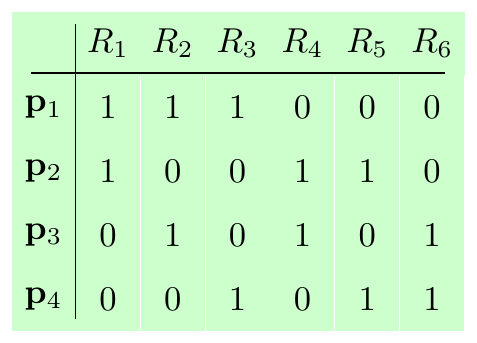}
\caption{An example of SFM $\mA^T$ for which $\UIDNC>\URLNC$ results in $\DIDNC>\DRLNC$.}\label{fig:sfm_2}
\end{figure}
Consider the SFM in Fig. \ref{fig:sfm_2}. Assume RLNC is applied, then $\URLNC=\wmax=2$ and $\DRLNC=2$. If IDNC is applied, because all four data packet conflict with each other, they must be sent separately. Thus, $\UIDNC=4$ and $\DIDNC=2.5$.
\end{Example}

The question that motivates our following work is whether there exist coding scheme(s) that can provide a block completion time between $\URLNC$ and $\UIDNC$ and an average packet decoding delay between $\DIDNC$ and $\DRLNC$. If one can find such schemes then moderate throughput-delay tradeoffs compared with IDNC and RLNC can be achieved. We will show that this is possible by packet partitioning.

\section{New Coding Framework}\label{sec:framework}
\subsection{Motivation}\label{sec:partition_motivation}

The initial motivation of this work is to investigate whether the decoding delay of RLNC can be reduced without sacrificing on throughput. RLNC suffers from large decoding delay mainly because it encodes all $K$ partially-received data packets in a block $\P_K$ together. Hence, one may infer that decoding delay could be reduced by partitioning  these $K$ data packets into several \emph{smaller generations} and applying RLNC to these small generations separately. By doing so, data packets in early small generations can be decoded sooner. The idea of applying small generations has been studied in the literature for traditional RLNC, e.g., \cite{heide_systematic_RLNC} for reducing decoding complexity and \cite{emina:joshi:isit2013} for reducing decoding delay, where throughput loss has been reported as inevitable. We now show that such partitioning does not generally work well for systematic RLNC either.

In this paper we use the term \emph{sub-generations} to denote smaller generations after partitioning of a partially-received data block $\P_K$ of original size $K$. A sub-generation, by classic definition of generation \cite{fragouli:lun:medard:pakzad:2007}, is a set of $g$ consecutive data packets, where $g$ is called the sub-generation size and ranges from 1 to $K$ here. According to this definition, sub-generations are generated by consecutively and evenly partitioning $K$ data packets in $\P_K$ into $M=\lceil K/g\rceil$ partitions. Denote by $\mG_m$ the $m$-th sub-generation, we have $\mG_m=\{\p_{(m-1)g+1},\cdots,\p_{mg}\}$. Its coded packet is:
\begin{equation}\label{eq:coded_packets}
\X^m=\sum_{\p_k\in\mG_m}\alpha_k^m\p_k
\end{equation}
where coefficients $\{\alpha_k^m\}$ are randomly chosen from an appropriate finite field $\F_q$.
Let $\wmax^m$ be the largest number of data packets in $\mG_m$ wanted by any one receiver across all receivers. $\mG_m$ thus requires a minimum of $\wmax^m$ coded transmissions. Then the minimum block completion time under partitioning with sub-generation size $g$, denoted by $U_g$, is calculated as:
\begin{equation}
U_g=\sum_{m=1}^{M}\wmax^m\label{eq:u_calculation}
\end{equation}

We further denote by $D_g$ the minimum average packet decoding delay under sub-generation size $g$. Its calculation follows \eqref{eq:d_definition}. We now demonstrate via an example that this classic partitioning can result in large $U_g$ and $D_g$ and hence is undesirable in terms of both throughput and delay.

\begin{Example}\label{exmp:bad_partition}
Consider the SFM given in Fig. \ref{fig:sfm}.  $K=8$ data packets can be partitioned into $M=2$ sub-generations under $g=4$: the first sub-generation is $\mG_1=\{\PP_1,\PP_2,\PP_3,\PP_4\}$ and the second sub-generations is $\mG_2=\{\PP_5,\PP_6,\PP_7,\PP_8\}$. Because $\wmax^1=3$ and $\wmax^2=4$, the minimum block completion time is $U_g=7$, which is much greater than both $\UIDNC$ and $\URLNC$. The minimum average packet decoding delay is $D_g=4.2$, which is also much greater than both $\DIDNC$ and $\DRLNC$ studied in Example \ref{exmp:sfm_idnc} and below \eqref{eq:d_rlnc}.
\end{Example}

However, by changing the content of each sub-generation, we can obtain better performance:
\begin{Example}\label{exmp:good_partition}
Consider the SFM given in Fig. \ref{fig:sfm}. We still partition the data packets into two sub-generations, but the first sub-generation is $\mG_1=\{\p_2,\p_3,\p_4,\p_7,\p_8\}$ and the second sub-generation is $\mG_2=\{\p_1,\p_5,\p_6\}$. The two sub-generations have $\wmax^1=\wmax^2=2$ and thus $U_g=4$, which is the same as $\URLNC$. However, since data packets in $\mG_1$ can be decoded after only 2 coded transmissions, $D_g$ is only 2.7, which is smaller than $\DRLNC=3.6$ calculated earlier.
\end{Example}

Example \ref{exmp:bad_partition} demonstrates that the classic partitioning does not provide convincing throughput and decoding delay performance. Whereas Example \ref{exmp:good_partition} shows that there may be better ways of partitioning. Explicitly, under classic partitioning, the minimum block completion time $U_g$ has the following property:

\begin{Lemma}
If data packets are partitioned into classic sub-generations, $U_g$ is lower bounded as:
\begin{equation}
U_g\geqslant\max(\wmax,M)
\end{equation}
\end{Lemma}
\begin{IEEEproof}
It is obvious that $U_g\geqslant\wmax$. Furthermore, because $\wmax^m\geqslant 1$ for $m\in[1,M]$, $U_g\geqslant M$ according to (\ref{eq:u_calculation}).
\end{IEEEproof}

This lemma indicates that, when a small sub-generation size $g$ is applied in an attempt to reduce decoding delay, the resulted $M$ and $U_g$ can be much larger than $\URLNC$ and even $\UIDNC$. A large $U_g$ implies large decoding delays for data packets sent in the last few sub-generations. This will in turn increase $D_g$, making such partitioning pointless in terms of both throughput and decoding delay. In Example \ref{exmp:bad_partition}, due to the fact that $U_g=7$, the decoding delay of data packets $\p_{5\sim8}$ is as large as 7.

In conclusion, following the classic partitioning we are unable to fill the performance gap between IDNC and RLNC. A better way of partitioning is needed and will be developed next, based on which the concept of sub-generations will be redefined and a new coding framework will be proposed.

\subsection{New Definitions and Coding Framework}

The partitioning in Example \ref{exmp:bad_partition} failed to reduce either $U_g$ or $D_g$ because all data packets in the same sub-generation, e.g., $\mG_2=\{\PP_5,\PP_6,\PP_7,\PP_8\}$, are jointly wanted by at least one receiver, yielding a minimum of $\wmax^2=4$ coded transmissions to complete this sub-generation. In contrast, as shown in Example \ref{exmp:good_partition}, sub-generations of $\mG_1=\{\p_2,\p_3,\p_4,\P_7,\p_8\}$ and $\mG_2=\{\p_1,\p_5,\p_6\}$ only requires $\wmax^1=\wmax^2=2$ coded transmissions because all receivers want at most 2 data packets from them.

These two examples motivate the key to a better partitioning, that is, \emph{to avoid as much as possible partitioning data packets that are jointly wanted by any receiver into the same sub-generation.} By doing so, $\{\wmax^m\}$ of the sub-generations are reduced. Imagine an extreme sub-generation in which any two data packets are not jointly wanted by any receiver. Then the broadcast of this sub-generation only requires one coded transmission. Such a sub-generation, recalling Definition \ref{def:coding_set} in Section \ref{sec:system_idnc}, is exactly an IDNC coding set.

The key step of partitioning then becomes clear:
\begin{Proposition}
Instead of partitioning packets in the packet set $\P_K$ based on their consecutive index, we partition a set of IDNC coding sets which together cover all $K$ data packets. In other words, we partition an IDNC solution.\footnote{Since the optimal IDNC solution $\S$ has the smallest cardinality ($\UIDNC$), it is the desired object for partitioning. Although we will develop the new coding framework based on the optimal IDNC solution, the proposed definitions, properties, and implementations are not restricted to the optimal IDNC solution. They can be applied to any IDNC solution satisfying Definition \ref{def:idnc_solution}, such as those heuristically found in \cite{yu:parastoo:neda:idnc2013}.}
\end{Proposition}

Accordingly, the concept of sub-generations is redefined:

\begin{Definition}
A sub-generation $\mG$ is a collection of IDNC coding sets in an IDNC solution.
\end{Definition}
The definition of sub-generation size is also changed:
\begin{Definition}
Sub-generation size $g$ is the number of coding sets in a sub-generation.
\end{Definition}

\begin{figure}
\centering
\includegraphics[width=0.85\linewidth]{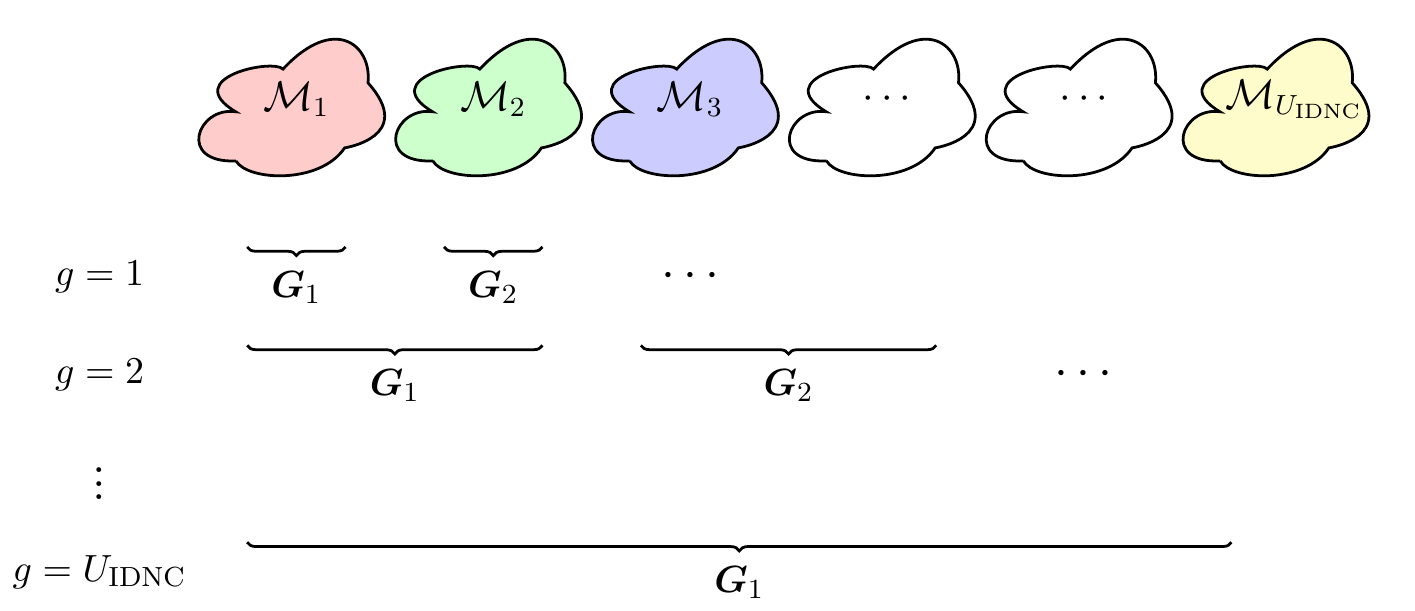}
\caption{The proposed coding framework with $g\in[1,\UIDNC]$.}
\label{fig:partition}
\end{figure}

The above definitions enable a new coding framework. Given the optimal IDNC solution $\S=\{\M_1,\cdots,\M_{\UIDNC}\}$, the $\UIDNC$ maximal coding sets are partitioned into $M=\lceil\UIDNC/g\rceil$ sub-generations, where  $g\in[1,\UIDNC]$. The simplest partitioning method is a consecutive one: $\mG_1=\{\M_1,\cdots,\M_g\}$, $\mG_2=\{\M_{g+1},\cdots, \M_{2g}\}$, $\cdots$.  A schematic of this partitioning is plotted in \figref{fig:partition}. More efficient partitioning algorithms will be developed in Section \ref{sec:implementations}.

For each sub-generation $\mG_m$, its coded packets are generated using \eqref{eq:coded_packets} as in classic partitioning. Linear independency among the coded packets are promoted by randomly choosing coding coefficients from a sufficiently large finite field $\F_q$. The field size $q$ can be reduced with decreasing sub-generation size $g$. In fact, $q=2$ when $g=1$, since all receivers want at most one data packet from a sub-generation of size $g=1$. The definition of $\wmax^m$ of a sub-generation $\mG_m$, where $m\in[1,M]$, is still the largest number of data packets in $\mG_m$ wanted by any one receiver across all receivers. Since $\mG_m$ needs a minimum of $\wmax^m$ coded transmissions, the relationship between $U_g$ and $\wmax^m$ is still as in \eqref{eq:u_calculation}. Below is an example of the proposed coding framework.

\begin{Example}
Consider the SFM in Fig. \ref{fig:sfm}, whose optimal IDNC solution is: $$\S=\{\{\p_2,\p_3,\p_7\},\{\p_4,\p_8\},\{\p_1,\p_5\},\{\p_1,\p_6\}\}$$ If $g=2$, there will be $M=\UIDNC/g=2$ sub-generations, where $\mG_1=\{\p_2,\p_3,\p_4,\p_7,\p_8\}$ and $\mG_2=\{\p_1,\p_5,\p_6\}$. This result is exactly the same as that in Example \ref{exmp:good_partition} and explains the success of the partitioning in Example \ref{exmp:good_partition} compared to the classic partitioning in Example \ref{exmp:bad_partition}.
\end{Example}

Interestingly, there are two extreme cases of this coding framework, taking place when $g=1$ and $g=\UIDNC$. When $g=1$, since $q=2$, coding within a sub-generation is through XOR and there are $\UIDNC$ such sub-generations. Hence, the coding scheme becomes IDNC with $U_1=\UIDNC$. When $g=\UIDNC$, there will be only one sub-generation, which contains all coding sets and thus all $K$ data packets. Hence, the coding scheme becomes RLNC with $U_{\UIDNC}=\URLNC$.

Therefore, the proposed coding framework successfully unifies the coding mechanisms of IDNC and RLNC. It also enables the coding schemes in the spectrum between IDNC and RLNC with $g\in[2,\UIDNC-1]$. In the next section, we will study throughput and decoding delay properties of the proposed coding framework under all values of sub-generation size $g$.

\section{Throughput and Decoding Delay Properties}\label{sec:properties}

The minimum block completion time $U_g$ is determined by $\{\wmax^m\}$ according to (\ref{eq:u_calculation}). $\{\wmax^m\}$ has the following property:
\begin{Lemma}
When sub-generation size satisfies $g\geqslant 2$, $\wmax^m\in[2,g]$ for all sub-generations $m\in[1,M]$.
\end{Lemma}
\begin{IEEEproof}
\begin{enumerate}[(1)]
\item $\mG_m$ is the superset of all $g$ IDNC coding sets in it. If all receivers want at most one data packet in $\mG_m$, i.e., if $\wmax^m=1$, then $\mG_m$ itself is an IDNC coding set.  This contradicts with the fact that the union of any coding sets in an IDNC solution is not a coding set according to Definition \ref{def:idnc_solution}. Thus $\wmax^m\geqslant2$;
\item If there is a receiver who wants $\wmax^m>g$ data packets in $\mG_m$, these $\wmax^m$ data packets conflict with each other and at least two of them must belong to the same coding set. This contradicts the IDNC constraint stated in Definition \ref{def:coding_set}. Thus, $\wmax^m\leqslant g$.
\end{enumerate}
\end{IEEEproof}

Then, by considering the relationship between $\wmax^m$ and $U_g$ in (\ref{eq:u_calculation}) and noting that $M=\lceil\UIDNC/g\rceil$, the above lemma yields an important corollary:
\begin{Corollary}\label{theo:U_bounds}
For all values of sub-generation size $g$, the minimum block completion time is bounded between $\URLNC$ and $\UIDNC$:
\begin{equation}
\forall g:~U_g\in[\URLNC,\UIDNC]
\end{equation}
where $U_g=\URLNC$ holds when $g=\UIDNC$ and $U_g=\UIDNC$ holds when $g=1,2$ (because $\wmax^m=2$ when $g=2$).
\end{Corollary}

Explicitly, $U_g$ approaches $\URLNC$ when the sub-generation size $g$ increases gradually from $1$ to $\UIDNC$ and approaches $\UIDNC$ the other way around.

The above lower and upper bounds on the minimum block completion time generally hold under the proposed coding framework. They are also useful benchmarks for the throughput performance of packet-level network coded wireless broadcast schemes in general. While $\URLNC$ is the lower bound for any such scheme, the upper bound $\UIDNC$ is an indicator of throughput efficiency, because any scheme requiring a minimum block completion time of greater than $\UIDNC$ can be treated as inefficient. One such inefficient scheme is the RLNC with classic partitioning in Example \ref{exmp:bad_partition}.

For the minimum average packet decoding delay, $D_g$, the situation is more complicated. It is clear that $D_g$ lies between $\DIDNC$ and $\DRLNC$. However, as we have discussed in Remark \ref{remark:rlnc_vs_idnc}, there is no guarantee that $\DIDNC<\DRLNC$. Hence, instead of $\DIDNC$ or $\DRLNC$, we propose an alternative upper bound on $D_g$ in terms of sub-generation size $g$:

\begin{Lemma}
The minimum average decoding delay under sub-generation size $g$ satisfies
\begin{equation}
D_g\leqslant \frac{g+\UIDNC}{2}\label{eq:D_bound}
\end{equation}
\end{Lemma}
\begin{IEEEproof}
Denote the largest decoding delay of data packets in $\mG_m$ by $D_g(m)$. It is equal to $\sum_{i=1}^{m}\wmax^i$. When $\{\wmax^i\}$ is maximized, i.e., when $\wmax^i=g$ for all $i\in[1,m]$, $D_g(m)$ is maximized with a value of $mg$. Since we can always send $\mG_m$ with more targeted receivers first, the largest $D_g$ happens when all $\mG_m$ have the same number of target receivers, denoted by $T$. Then, as a variation of (\ref{eq:d_definition}), the largest $D_g$ is calculated as:
\begin{equation}
D_g=\frac{1}{MT}\sum_{m=1}^MTmg=\frac{g(1+M)}{2}=\frac{g+\UIDNC}{2}
\end{equation}
\end{IEEEproof}

This bound also justifies the application of the systematic transmission phase: At the beginning of the broadcast of $K_T$ data packets, we have an all-one SFM $\mA$ of size $K_T\times K_T$. Thus, while $\UIDNC=\URLNC=K_T$, (\ref{eq:D_bound}) indicates that using $g=1$ offers the smallest packet decoding delay, which requires all $K_T$ data packets to be sent separately uncoded.

In summary, in this section we showed that throughput and decoding delay performance of all coding schemes in the proposed coding framework is well bounded between that of IDNC and RLNC. Therefore, our coding framework is a general one with IDNC and RLNC identified as two extreme cases with $g=1$ and $g = \UIDNC$, respectively. It successfully fills the performance gap between IDNC and RLNC and enables a series of coding schemes with a range of more balanced throughput and decoding delay performance. In the next section, we will turn to implementations of the proposed coding framework to further improve its throughput and decoding delay performance.

\section{Implementations}\label{sec:implementations}
In the last section, we showed that throughput and decoding delay performance of the proposed coding framework is well bounded between IDNC and RLNC for all sub-generation sizes $g\in[1,\UIDNC]$. Although we cannot further improve the performance of IDNC and RLNC, we can do so for the coding schemes with $g\in[2,\UIDNC-1]$.

After the systematic transmission phase, there are two steps in the coded transmission phase: 1) partitioning IDNC coding sets into sub-generations; and 2) broadcasting these sub-generations following a transmission strategy. We will first optimize these two steps and then further explore the potentials of this coding framework.

\subsection{Partitioning Algorithms}

We denote by $T(u)$ the number of targeted receivers of a coding set $\M_u$ in the optimal IDNC solution $\S=$ $\{\M_1,$ $\M_2,$ $\cdots,$ $\M_{\UIDNC}\}$, where:
\begin{equation}
T(u)=\sum_{\PP_k\in\M_u}T_k
\end{equation}

Without loss of generality we also assume that $T(1)\geqslant T(2)\geqslant \cdots \geqslant T(\UIDNC)$, i.e., $\M_1$ is wanted by the most receivers, followed by $\M_2$, and so on.

The simplest algorithm is to consecutively partition the $\UIDNC$ coding sets. That is, $\mG_1=\{\M_1,\cdots,\M_g\}$, $\mG_2=\{\M_{g+1},\cdots,\M_{2g}\}$, $\cdots$. This algorithm is referred to as \emph{Direct partitioning} (DP).

DP offers good decoding delay performance, because coding sets with more targeted receivers are partitioned into earlier sub-generations and are broadcast earlier. However, $\wmax^m$ may not be uniform across coding sets in $\mG_m$. DP overlooks this fact and thus does not necessarily minimize $\{\wmax^m\}$ and $U_g$. Below is an example.

\begin{Example}
\begin{figure}
\centering
\includegraphics[width=0.4\linewidth]{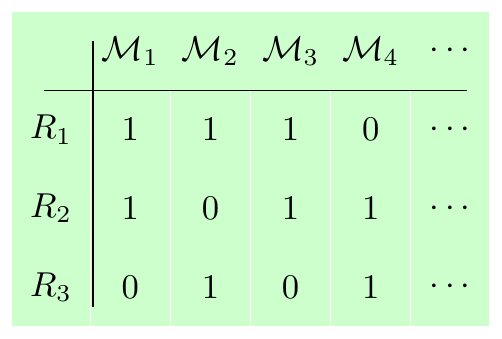}
\caption{An example of smartly reducing $\wmax^1$ when $g=3$.}\label{fig:algorithmc_partition}
\end{figure}

\emph{In the matrix in Fig. \ref{fig:algorithmc_partition}, an entry of one at row $n$ and column $i$ means receiver $R_n$ wants one data packet in coding set $\M_i$. Suppose the sub-generation size is $g=3$. Then, according to DP, we have $\mG_1=\{\M_1,\M_2,\M_3\}$, and thus $\wmax^1=3$. However, if we use $\mG_1=\{\M_1,\M_2,\M_4\}$, we have $\wmax^1=2$.}
\end{Example}

Hence, to strike a balance between throughput and decoding delay, we design a heuristic algorithm which aims at 1) reducing $U_g$ by exploiting the opportunities of reducing $\wmax^m$ for all $m$, and 2) preserving low decoding delay by partitioning the most wanted coding sets into earlier sub-generations. This algorithm, which we refer to as \emph{Smart partitioning} (SP), fills sub-generations sequentially, i.e., the $g$ coding sets for $\mG_1$ are first determined, followed by the $g$ coding sets for $\mG_2$, etc. Details of this algorithm are presented in Algorithm \ref{alg:partition}.
\begin{algorithm}
\caption{Smart partitioning (SP)}
\label{alg:partition}
\begin{algorithmic}[1]
\STATE initialize: an IDNC solution $\S$, and $M$ empty sub-generations, $\mG_1,\cdots,\mG_M$;
\STATE sort the coding sets in $\S$ in a descending order in terms of their number of targeted  receivers;
\FOR {$m=1:M$}
\FOR {$i=1:g$}
\STATE find the coding sets in $\S$ that do not increase $\wmax^m$ by one. Denote the collection of such coding sets by $\S'$;
\IF {$\S'$ is not empty}
\STATE add the coding set in $\S'$ with the smallest index (\emph{and thus wanted by most receivers}) to $\mG_m$;
\ELSE
\STATE add the coding set in $\S$ with the smallest index to $\mG_m$;
\ENDIF
\STATE remove the chosen coding set from $\S$;
\IF {$\S$ is empty}
\STATE terminate the algorithm;
\ENDIF
\ENDFOR
\ENDFOR
\end{algorithmic}
\end{algorithm}

Compared with DP, SP costs light extra computations. However, as will be numerically compared later, the performance of SP is better than DP in a wide range of system settings. The reason why SP outperforms DP in terms of decoding delay is that, by reducing $U_g$ in SP we also reduce the worst packet decoding delay, which will in turn reduce the average packet decoding delay.

With sub-generations generated, we are ready to broadcast them through erasure-prone channels.

\subsection{Coded Transmission Strategies}\label{sec:trans_schemes}
In this subsection, we present two coded transmission strategies, called Sequential and Semi-online, respectively. They are superior to each other in different aspects.
\subsubsection{Sequential strategy}

~

Given all $M$ sub-generations, the simplest strategy for the coded transmission phase is to segment this phase into $M$ rounds. In each round, coded packets of a sub-generation are broadcast until all its targeted receivers have decoded it and informed the sender. We name this strategy \emph{Sequential}.

\begin{figure}
\centering
\includegraphics[width=0.85\linewidth]{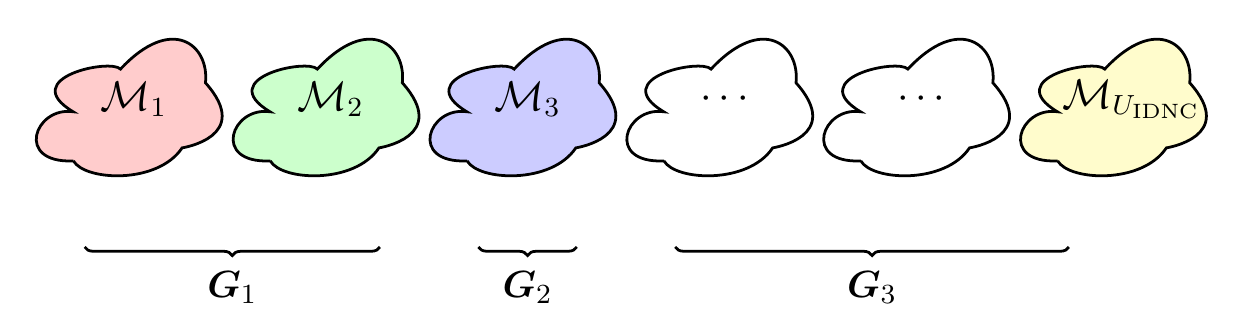}
\caption{Changing the sub-generation size $g$ during the broadcast using Sequential strategy.}
\label{fig:change_g}
\end{figure}

In Sequential strategy, the decoding of different sub-generations are independent of each other. This property leads to the primary advantage of this strategy, namely, tunable throughput and decoding delay even within the broadcast of a data block, because we can apply different sub-generation sizes in different rounds without affecting the decoding of sub-generations in other rounds. For example, when decoding delay becomes the primary concern, the system can easily switch to IDNC by setting $g=1$ in all remaining rounds. Such switching was impossible because RLNC generally applies a larger field size than IDNC. Another example is schematically shown in \figref{fig:change_g}. In this example, there are three rounds, with sub-generation sizes of 2, 1, and $\UIDNC-3$, respectively.

Sequential strategy requires at most $M$ ACK feedback from every receiver. The sub-generation size can also be adjusted to fit the system's specific feedback frequency if there is any.

The main disadvantage of this strategy is its large decoding delay, because the broadcast of a sub-generation $\mG_m$ cannot be started until the broadcast of
$\mG_{m-1}$ is completed. Even if only one targeted receiver of $\mG_{m-1}$
experiences a bad channel, the round for $\mG_{m-1}$ must continue and thus all targeted receivers of $\mG_m$ have to wait. To overcome this drawback, we propose another coded transmission strategy called \emph{Semi-online} strategy.

\subsubsection{Semi-online strategy}
~

In Semi-online strategy, the coded transmission phase is also segmented into rounds, which is, though, different from the rounds in Sequential strategy. Here in each round, all
$M$ sub-generations are broadcast, where for $\mG_m$, $\wmax^m$ coded packets
are broadcast. Thus in each round, there are $U_g=\sum_{m=1}^M\wmax^m$
coded transmissions. After each round, the sender collects feedback from the
receivers about how many more coded packets of each sub-generation they still want. $\{\wmax^m\}$ is thus updated accordingly before the next round starts. Below is a simple example.
\begin{Example}
Assume there are three receivers, $R_1$ to $R_3$. They want two, three, and four data packets from sub-generation $\mG_1$, respectively. Therefore, $\wmax^1=4$ coded packets of $\mG_1$ are broadcast in the first round, along with coded packets of other sub-generations. Assume that after the first round, $R_1$ to $R_3$ have received two, one, and three coded packets of $\mG_1$, respectively. Then they still want zero, two, and one coded packets to decode $\mG_1$, respectively. $\wmax^1$ is thus updated to two. In the second round, two coded packets of $\mG_1$ will be broadcast, along with coded packets of other sub-generations.
\end{Example}

Semi-online strategy overcomes the main drawback of Sequential
strategy, and thus significantly improves the decoding delay performance.
On the other hand, from a statistical point of view, Semi-online strategy has the same throughput performance as Sequential strategy, since we only ``swap'' the coded transmissions for each sub-generation.

The main drawback of Semi-online strategy is that throughput and decoding delay performance is no longer tunable. Moreover, since the number of rounds in Semi-online strategy increases with decreasing channel quality, the amount of feedback in this strategy cannot be predetermined. However, as will be presented in Section \ref{sec:merge}, Semi-online strategy
offers the opportunity to algorithmically merge sub-generations together, which can further improve both throughput and decoding delay performance.

Therefore, there is no clear winner between Sequential and Semi-online strategies. Which one to adopt depends on the application. In the next subsection, we will discuss the concept and impacts of \emph{packet diversity} on these two strategies.

\subsection{Packet Diversity}

The diversity of a data packet in the proposed coding framework is defined as follows:
\begin{Definition}
The diversity of a data packet is the number of sub-generations in which it appears.
\end{Definition}

In the proposed coding framework, data packets might have diversities of greater than one because the sub-generations are generated from maximal IDNC coding sets, whose intersections are usually not empty. Our motivation of studying packet diversity is the fact that, we could possibly reduce $\wmax^m$ of a sub-generation $\mG_m$ by removing a data packet from it. Below is an example.
\begin{Example}
Assume that there are two sub-generations. The first sub-generation $\mG_1$ contains $\{\p_1$,$\p_2$,$\p_3\}$ and some other data packets. The second sub-generation $\mG_2$ contains $\p_3$ and some other data packets. Here data packet $\p_3$ has a diversity of 2. Also assume that receiver $R_1$ wants $\{\p_1$,$\p_2$,$\p_3\}$, which means $\wmax^1$ is at least 3. Now let us remove $\p_3$ from $\mG_1$, which reduces the diversity of $\p_3$ to 1. The targeted receivers of $\p_3$ can still decode it from $\mG_2$, while $\wmax^1$ can be possibly reduced to 2, which reduces the minimum block completion time $U_g$ by 1.
\end{Example}

Inspired by this fact, we investigate the benefits and problems that a packet diversity of greater than one brings to Sequential and Semi-online coded transmission strategies, and then decide whether it should be reduced to one or not.

\subsubsection{Packet diversity in Sequential strategy}
~

A packet diversity of greater than one is redundant in Sequential strategy regardless of the sub-generation size. Suppose data packet $\p_1$ is included in $\mG_1$ and $\mG_2$. By the end of the round for $\mG_1$, all receivers who want $\p_1$ will have received it, indicating that $\p_1$ does not need to be included in $\mG_2$.

\subsubsection{Packer diversity in Semi-online strategy}
~

Unlike Sequential strategy, the impacts of packet diversity in Semi-online strategy is much more complex. Since all data packets are broadcast in every round, a higher packet diversity can be translated into a higher probability of being received and decoded, and thus reduces the number of coded transmissions in the next round. However, a high packet diversity incurs complicated coding and decoding decision makings. This drawback can be illustrated by the following example.

\begin{Example}
Assume that receiver $R_1$ wants $\{\p_1,\p_2\}$ from $\mG_1$, and wants $\{\p_2,\p_3\}$ from $\mG_2$. Here $\p_2$ has a diversity of 2. Imagine a case that, after the first coded transmission round, $R_1$ has received one coded packet of $\mG_1$ and one coded packet of $\mG_2$. In the next round, $R_1$ only needs one coded packet of either $\mG_1$ or $\mG_2$ to decode all three data packets. Thus the sender needs to decide to send a coded packet of $\mG_1$ or $\mG_2$ or both. This case is referred to as Case-1. Imagine another case that, after the first coded transmission round, $R_1$ has received one coded packet of $\mG_1$ and two coded packets of $\mG_2$. $R_1$ can thus directly decode $\{\p_2,\p_3\}$ from $\mG_2$. After that, $R_1$ can substitute $\p_2$ into the received coded packet of $\mG_1$ to decode $\p_1$. This case is referred to as Case-2.
\end{Example}

When Case-1 in the above example is extended to all receivers, decision making by the sender  will become very complicated. When Case-2 in the above example is extended to all sub-generations, decoding by the receiver will become complicated, because once it has decoded some data packets from a sub-generation, it has to look up all other sub-generations for more decoding opportunities. The only exception happens when $g=1$, i.e., IDNC. In this case, since there is no linear equations to solve, such search is unnecessary.

In conclusion, we suggest to reduce the diversities of all data packets to one for both Sequential and Semi-online strategies unless $g=1$. This reduction is applied to the optimal IDNC solution by removing data packets from a maximal coding set if they have already been covered by the previous coding sets. The resulted solution is denoted by $\underline{\S}$ and also has a length of $\UIDNC$. Any of the partitioning algorithms discussed before can then be applied to $\underline{\S}$ before the coded transmission phase.

It is noted that our theoretical analysis and partitioning algorithms are not affected by the diversity reduction. Throughput properties are not affected because both $\UIDNC$ and $\URLNC$ remain the same and the relationship between $U_g$ and $\{\wmax^m\}$ always holds. Decoding delay properties are not affected because, by its definition in \eqref{eq:d_definition}, we only consider the first time slot that a data packet can be decoded. Its reception in later time slots due to diversity is not considered. Partitioning algorithms are not affected because they work for any valid IDNC solution.

In this subsection, we reduce packet diversities primarily for reducing $\{\wmax^m\}$ at the beginning of the coded transmission phase. In the next subsection, we will introduce an operation called \emph{sub-generation merging}, which could reduce $\{\wmax^m\}$ in the second and further coded transmission rounds in the Semi-online strategy.

\subsection{Sub-generation merging}\label{sec:merge}

Sub-generation merging is an operation that can only be applied under Semi-online strategy. We use a simple example to introduce it.

\begin{figure}[t]
\centering
\subfigure[before]{\begin{tabular}{c|c|c}
~ & $\mG_1$ & $\mG_2$\\\hline
$R_1$ & 2 & 0\\\hline
$R_2$ & 0 & 3
\end{tabular}}\hspace{50pt}
\subfigure[After]{\begin{tabular}{c|c}
~ & $\mG_1 \cup \mG_2$\\\hline
$R_1$ & 2\\\hline
$R_2$ & 3
\end{tabular}}
\caption{Number of coded packets wanted by the receivers before and after sub-generation merging.}\label{fig:dof_merge}
\end{figure}

\begin{Example}
Imagine that after a Semi-online coded transmission round, there are two uncompleted sub-generations and two receivers. As shown in Fig. \ref{fig:dof_merge}, $R_1$ still wants two coded packets of $\mG_1$ and $R_2$ still wants three coded packets of $\mG_2$. In this case, $\wmax^1=2$ and $\wmax^2=3$, thus $U_g=5$.
$R_1$ can decode after two coded transmissions, and $R_2$ can decode after five coded transmissions. Alternatively, let us merge $\mG_1$ and $\mG_2$ together, that is, combine data packets in $\mG_1$ and $\mG_2$ together to form a new sub-generation $\mG'$. $R_1$ wants two and $R_2$ wants three coded packets of $\mG'$, respectively. Thus $\wmax'=3$ and $U_g$ is reduced from 5 to 3, i.e., throughput is improved. Moreover, while $R_1$ can still decode after two coded transmissions, $R_2$ can decode after only three coded transmissions rather than five, thus decoding delay is also reduced.
\end{Example}

In this example, the two sub-generations $\mG_1$ and $\mG_2$ are not jointly wanted by any receiver. Compared with the definition of non-conflicting data packets in Section \ref{sec:system}, we define such sub-generations as \emph{non-conflicting sub-generations}. It is straightforward that non-conflicting sub-generations can be merged together so that both the throughput and decoding delay performance can be improved. The way of deciding which sub-generations to merge together is the same as finding an IDNC solution from its graph representation. Since the number of sub-generations is small, such decision making is not computationally expensive and can be heuristically found using the methods introduced in \cite{yu:parastoo:neda:idnc2013}.

\begin{figure}
\centering
\includegraphics[width=0.85\linewidth]{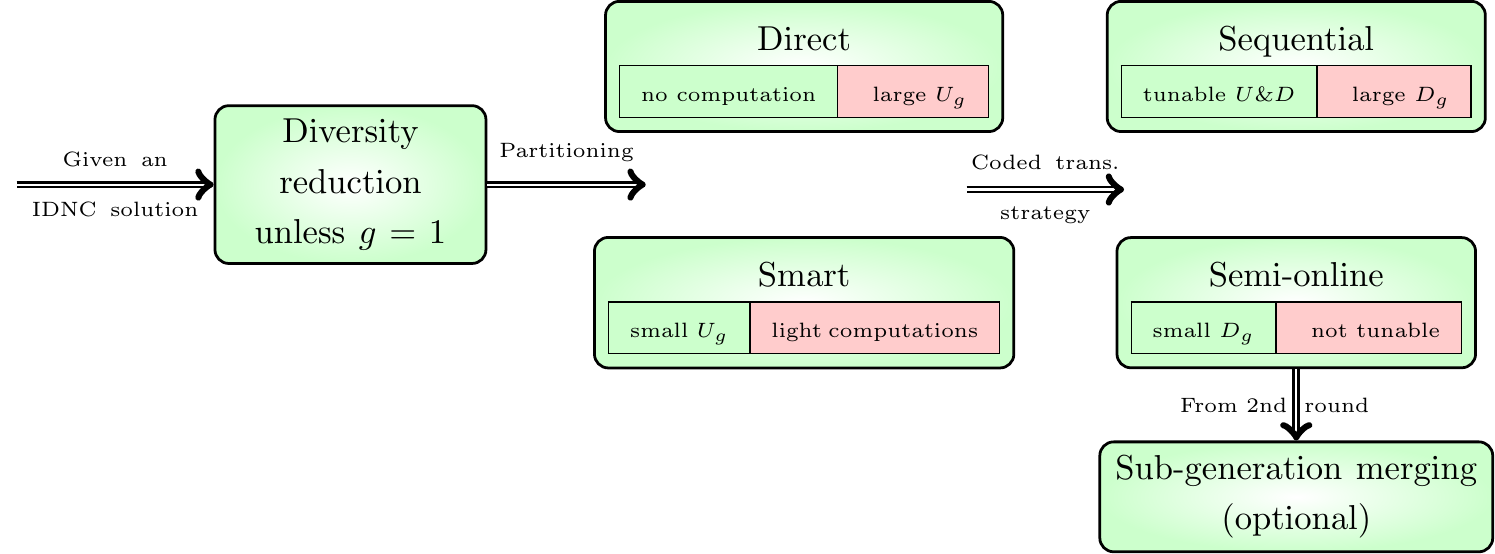}
\caption{A flow chart of the proposed implementations in \secref{sec:implementations} with their advantages and drawbacks.}
\label{fig:implementations}
\end{figure}

A flow chart is presented in \figref{fig:implementations} to summarize proposed implementations of our coding framework.

\begin{Remark}
The proposed coding framework and its implementations can be easily adapted to existing linear network coding systems, because at its core, the proposed coding framework generates linear combinations of the data packets as coded packets. The only modification is that the sender requires the receivers to provide feedback by the last coded transmission of each round.
\end{Remark}

\section{Simulations}\label{sec:simulations}
Four sets of simulations are carried out in this section. In the first simulation, we numerically demonstrate the well-bounded throughput and decoding delay performance of the proposed coding schemes compared with IDNC and RLNC. In the second to fourth simulations, we verify the effectiveness of the proposed implementations, including partitioning algorithms, coded transmission strategies, and sub-generation merging, respectively.

We simulate broadcast of $K_T=20$ data packets to $N_T=[5,50]$ receivers. Wireless channels between the sender and the receivers are subject to i.i.d. memoryless packet erasures with a probability of $P_e=0.2$. Since the first two simulations are concerned with the minimum block completion time and the minimum average packet decoding delay, packet erasures in the coded transmission phase are not considered in these two simulations, but will be incorporated in the third and fourth simulations.

\subsection{Achievable Throughput-delay Tradeoffs}
\begin{figure}
\centering
\includegraphics[width=0.95\linewidth]{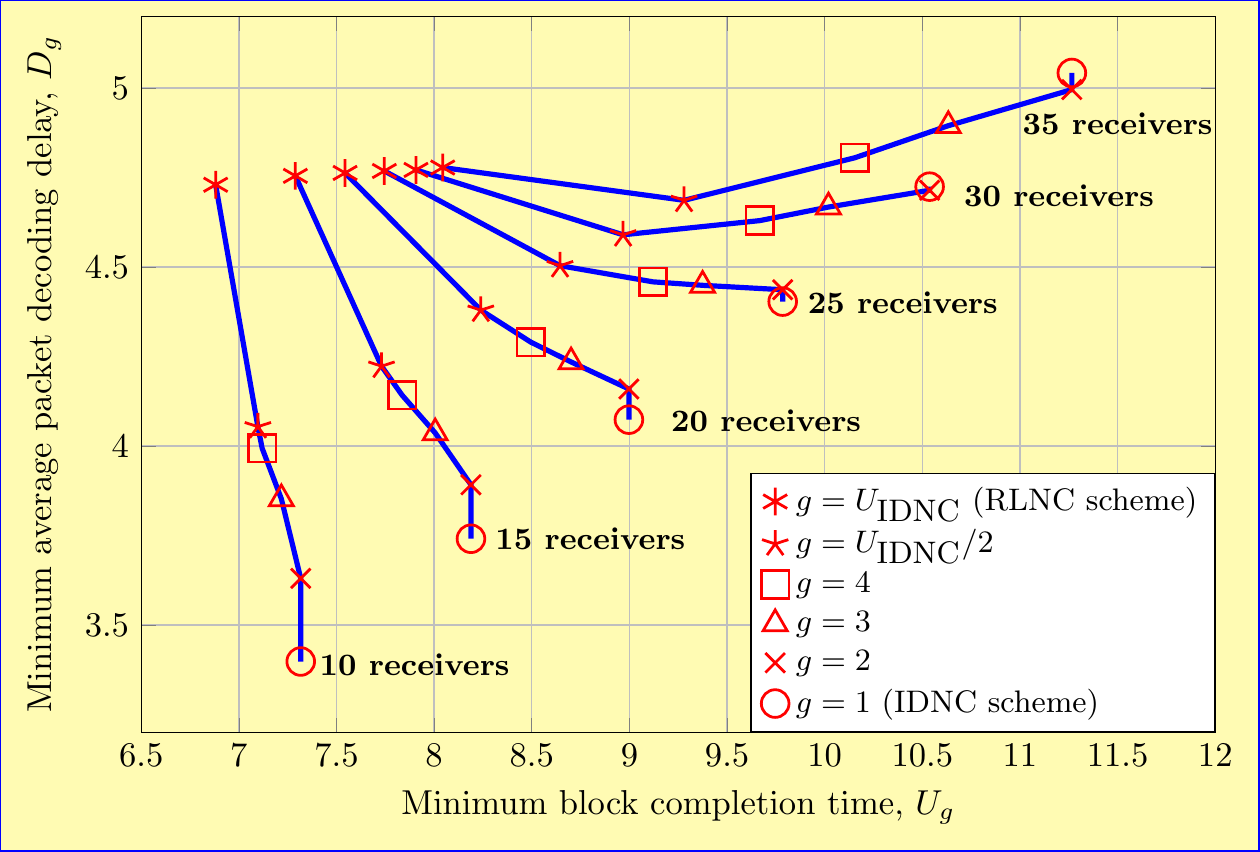}
\caption{Achievable throughput-delay tradeoffs of the proposed coding schemes with $K_T=20$ data packets.}
\label{fig:tradeoff}
\end{figure}

In this simulation, we evaluate the block completion time ($U_g$) and decoding delay ($D_g$) performance of the proposed coding schemes with various values of sub-generation size $g$, including $g=1$ (IDNC scheme), $g = 2, 3, 4, \UIDNC/2$, and $g = \UIDNC$ (RLNC scheme). Direct partitioning is applied. The results are plotted in \figref{fig:tradeoff}. As we can see, throughput and decoding delay performance of the coding schemes with $g\in[1,\UIDNC]$ is well bounded between the performance of IDNC and RLNC. They fill the performance gap between IDNC and RLNC, and thus offer moderate throughput-delay tradeoffs compared with IDNC and RLNC.

Another observation is that the throughput performance under $g=2$ is always the same as under $g=1$, i.e., $U_2=U_1=\UIDNC$. This result matches our upper bound on $U_g$ in Corollary \ref{theo:U_bounds}. On the other hand, their decoding delay performance is generally different due to their different sub-generation sizes.

\figref{fig:tradeoff} also provides an example where RLNC outperforms IDNC in terms of both throughput and decoding delay performance. It takes place when $\UIDNC$ becomes much larger than $\URLNC$ (11.3 and 8, respectively, at $N_T=35$), which means the worst packet decoding delay of IDNC is much larger than RLNC.

\subsection{Partitioning Algorithms}

\begin{figure*}
\centering
\subfigure[Throughput]{\includegraphics[height=0.37\linewidth]{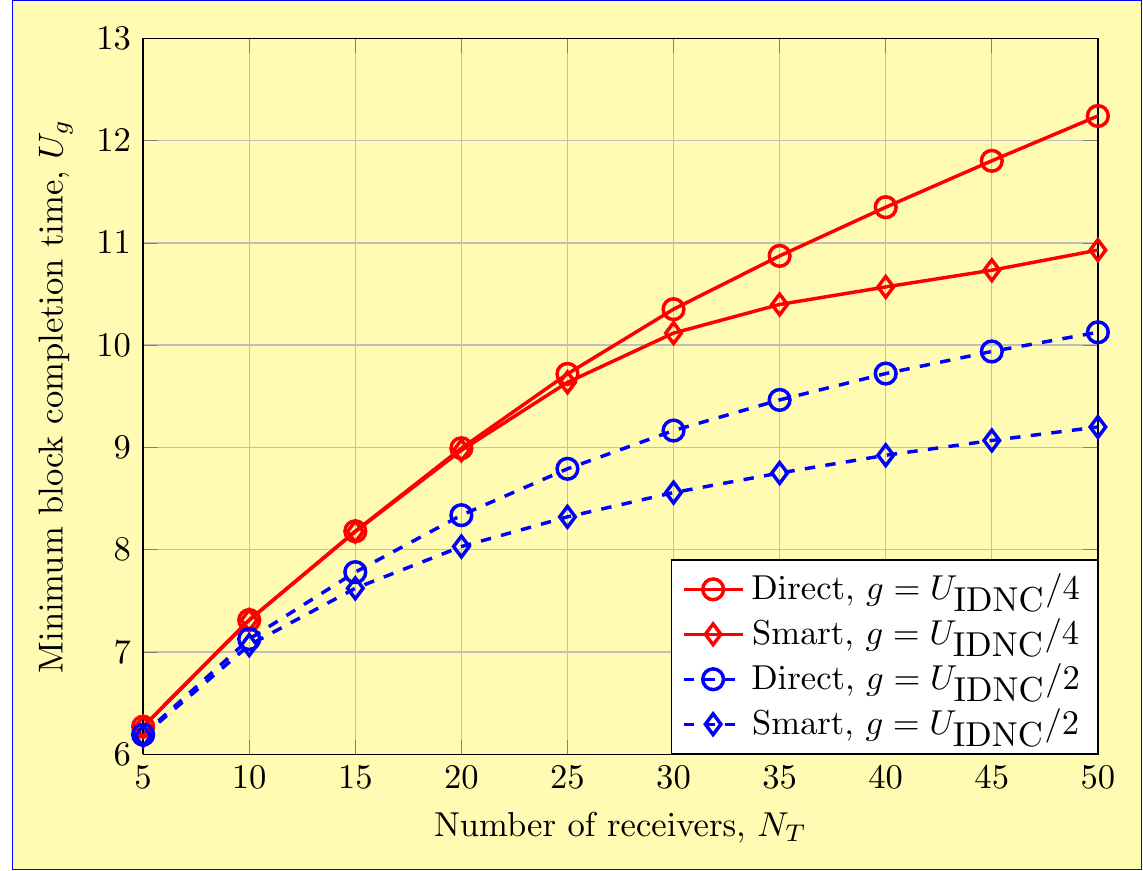}}
\subfigure[Decoding delay]{\includegraphics[height=0.37\linewidth]{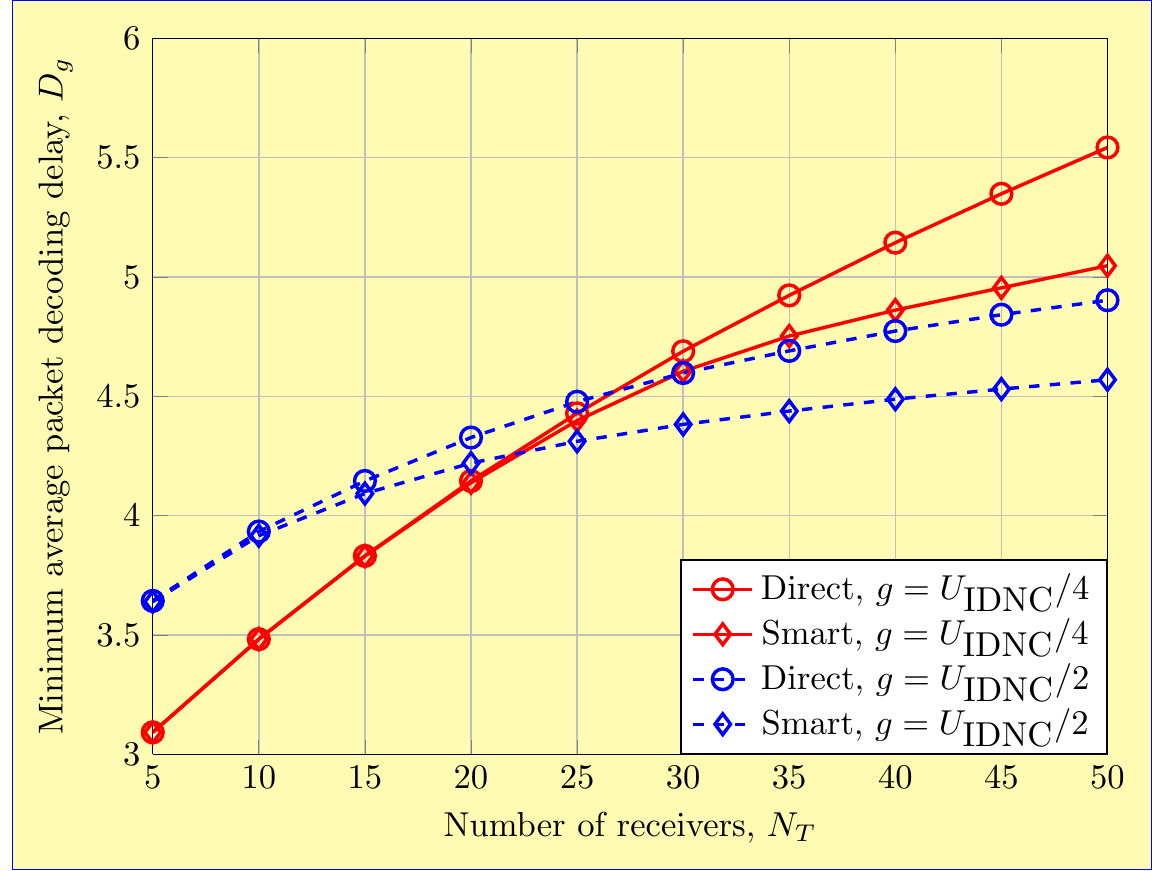}}
\caption{Performance comparisons between Direct and Smart partitionings.}\label{sim:partition}
\end{figure*}

In this simulation, we evaluate the Direct and Smart partitioning algorithms. We apply two sub-generation sizes, $g=\UIDNC/4$ and $g=\UIDNC/2$. Simulation results are shown in  Fig. \ref{sim:partition}. From this figure, we observe that the performance of Smart partitioning is equal to or better than Direct partitioning for all values of $N_T$ and for both values of sub-generation size $g$.

\subsection{Coded Transmission Strategies}

\begin{figure*}
\centering
\subfigure[Throughput]{\includegraphics[height=0.37\linewidth]{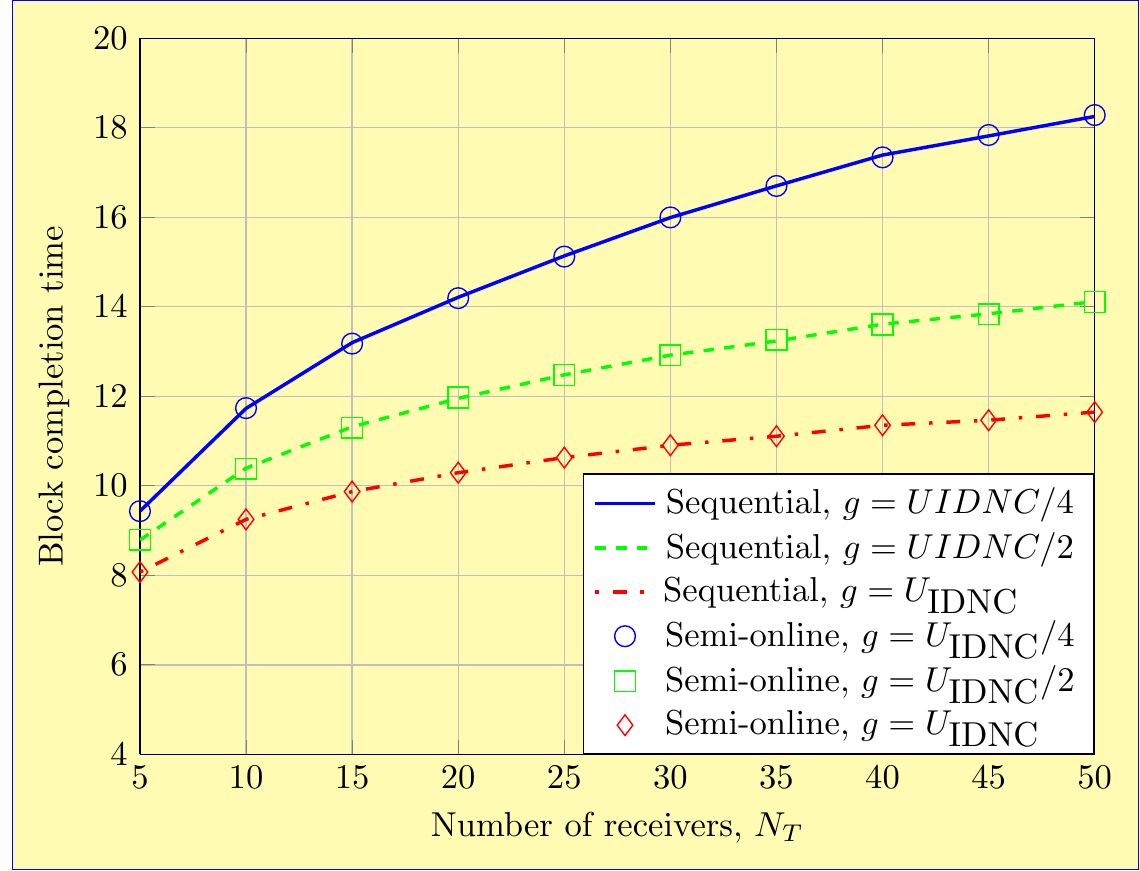}}
\subfigure[Decoding
delay]{\includegraphics[height=0.37\linewidth]{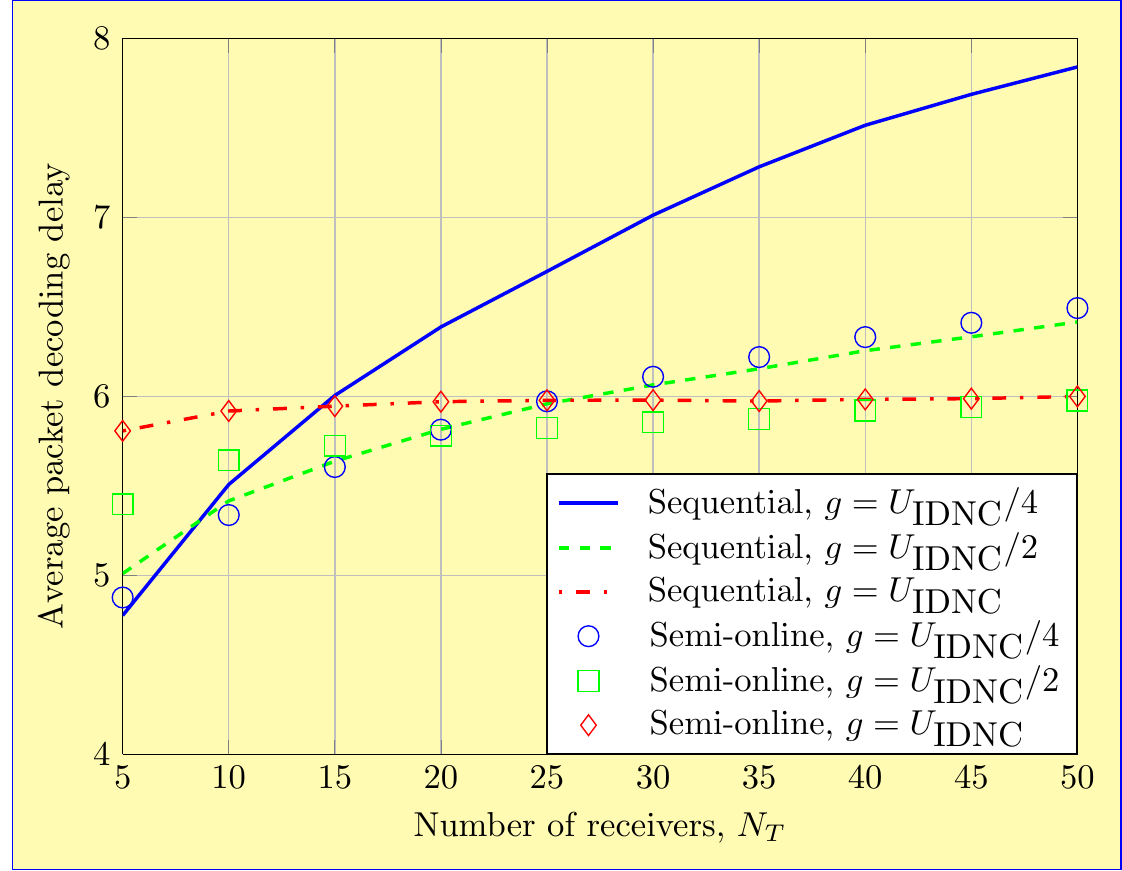}\label{sim:delay_straight_semi}}
\caption{Performance comparisons between Sequential and Semi-online coded transmission strategies.}\label{sim:straight_vs_Semi}
\end{figure*}

In this simulation, we evaluate Sequential and Semi-online coded transmission strategies. We use Smart partitioning and apply three sub-generation sizes, $g=\UIDNC/3$, $g=\UIDNC/2$ and $g=\UIDNC$. Simulation results are shown in  Fig. \ref{sim:straight_vs_Semi}. Our first observation is that the two strategies always share the same throughput performance. This result matches our statistical claim in Section \ref{sec:trans_schemes}. The second observation is that the decoding delay performance of the Semi-online strategy successfully outperforms the Sequential strategy when $g<\UIDNC$. When $g=\UIDNC$, since there is only one sub-generation, both strategies become equivalent to RLNC and thus have the same decoding delay performance.

\subsection{Sub-generation Merging}

\begin{figure*}
\centering
\subfigure[Throughput]{\includegraphics[height=0.37\linewidth]{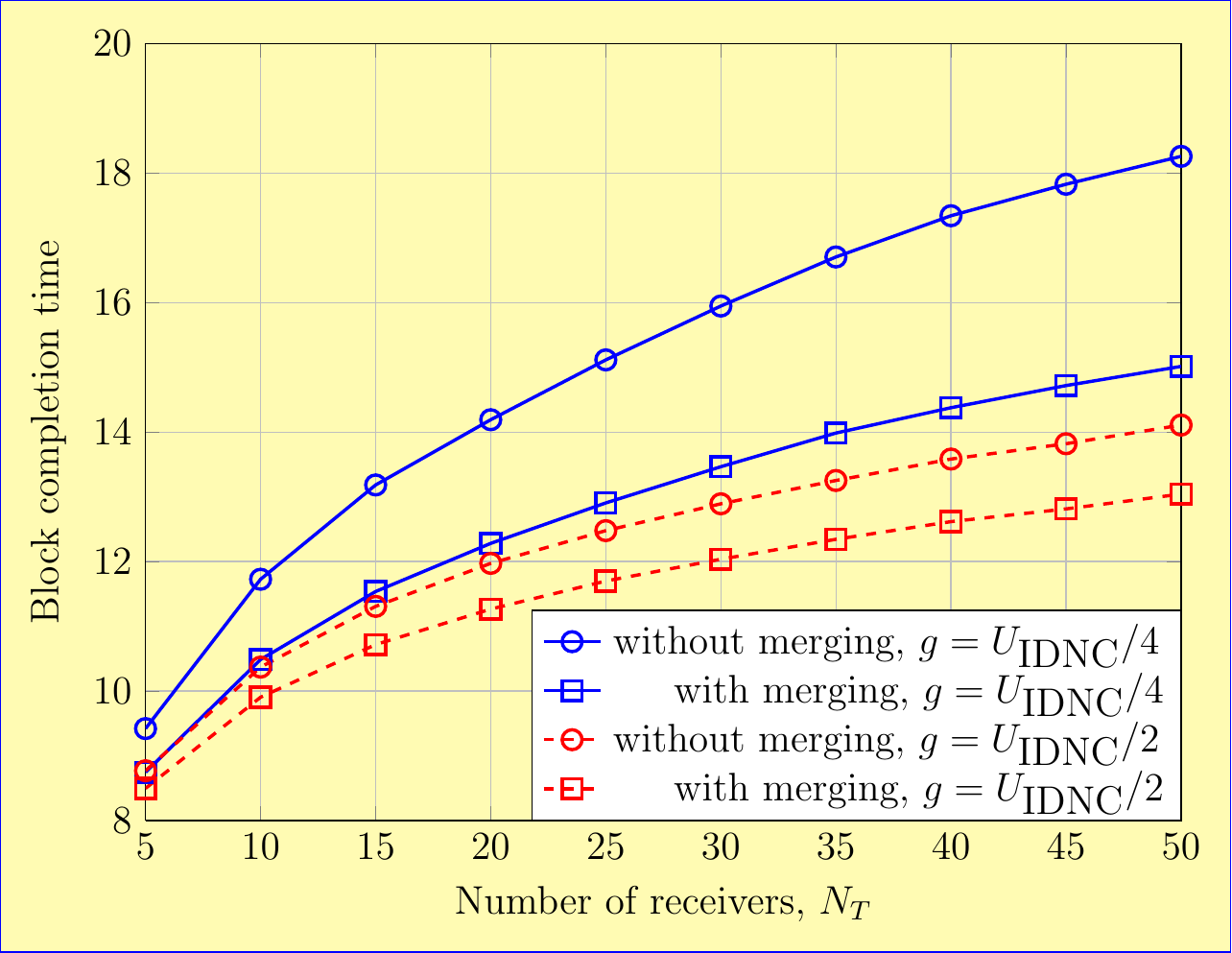}}
\subfigure[Decoding delay]{\includegraphics[height=0.37\linewidth]{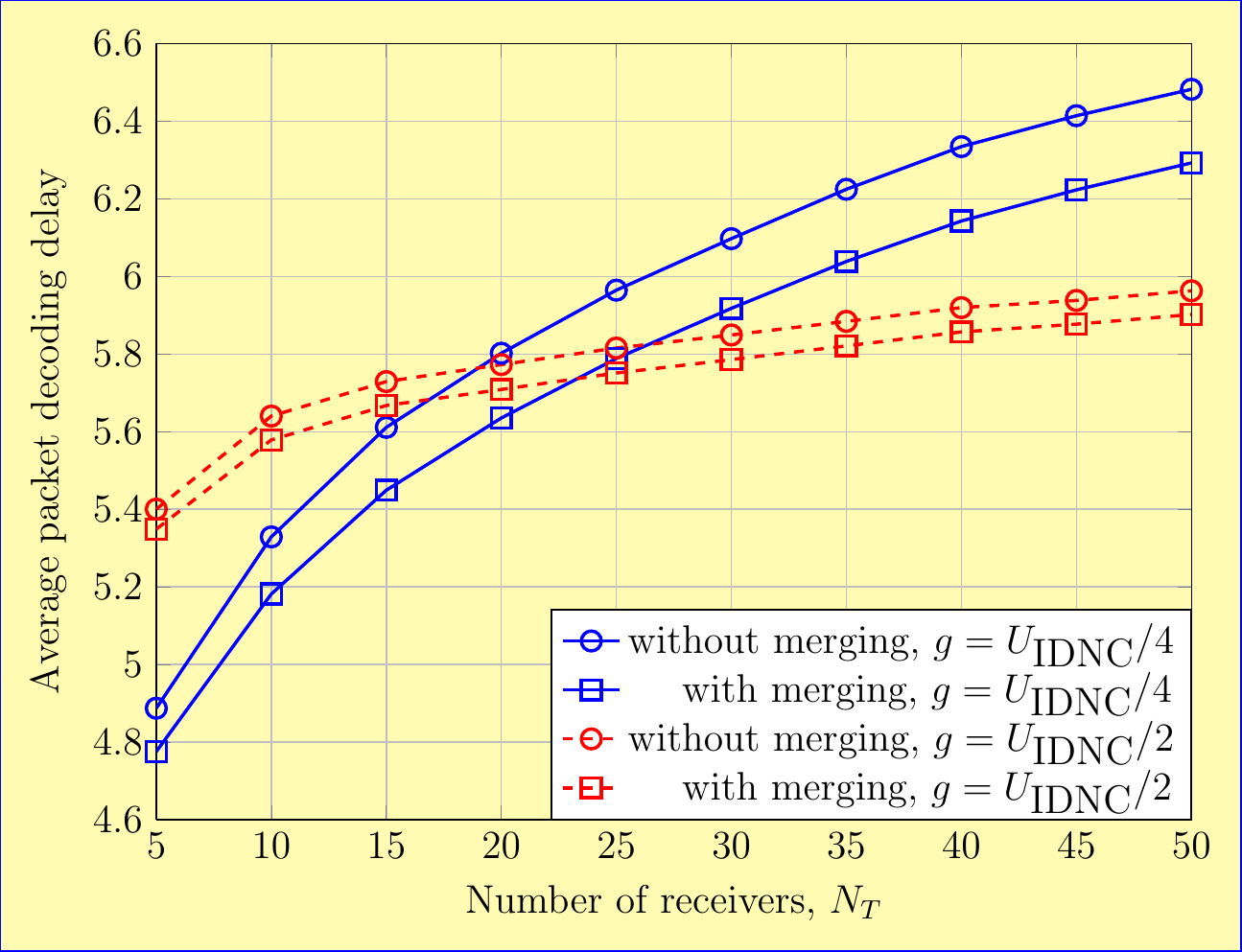}}
\caption{Performance comparisons between Semi-online strategy with and without sub-generation merging.}\label{sim:merge}
\end{figure*}

In this simulation, we compare the performance of Semi-online strategy with and without sub-generation merging. The results are shown in Fig. \ref{sim:merge}, from which it is clear that sub-generation merging improves both throughput and decoding delay under all parameter settings.

At a high level, our simulations show that throughput and decoding delay performance do not have to be improved by trading each other off, but they can have coordination in our coding framework. By applying proposed implementations and choosing a proper sub-generation size, a large range of throughput-delay tradeoffs can be achieved. The best configuration in terms of performance is Smart partitioning combined with Semi-online coded transmission strategy with sub-generation merging.

\section{Conclusion}

For wireless network coded broadcast, we showed that it is possible to build upon an IDNC solution to obtain a series of more general linear network coded solutions with varying throughput and decoding delay performance, as well as varying implementation complexity and feedback frequency. The core of our work was introducing a novel way of partitioning a partially-received data block into sub-generations based on the coding sets in a given IDNC solution. Consequently, when an IDNC solution is used unaltered for coded transmissions, we are at one end of the spectrum, namely IDNC with sub-generation size $g =1$. When all IDNC coding sets are combined, we reach the other end of the spectrum, namely RLNC with sub-generation size $g =\UIDNC$.

The primary advantage of our coding framework is that the throughput and decoding delay of all intermediate coding schemes with different sub-generation sizes are guaranteed to lie between those of IDNC and RLNC. With this crucial advantage, we showed that we are able to focus on further improvement of throughput and delay for intermediate coding schemes by more advanced partitioning and transmission strategies such as Smart partitioning combined with Semi-online transmission strategy and sub-generation merging, or even Smart partitioning with Sequential transmission strategy that enables in-block switching between IDNC and RLNC for performance adaptation.

The proposed coding framework and its implementations have the potential to be adapted to other linear network coding systems such as multi-hop systems.

\end{document}